\documentclass[%
 reprint,
 amsmath,amssymb,
 aps,nofootinbib,   
]{revtex4-1}
\usepackage{bm}
\PassOptionsToPackage{linktocpage}{hyperref}
\usepackage[hyperindex,breaklinks]{hyperref}
\usepackage{enumitem}
\usepackage{slashed}

\renewcommand{\theta}{\vartheta}

\usepackage{array}
\usepackage{mathtools}

\usepackage{etoolbox}

\begin{document} 

\title{ A String Theoretic Derivation of Gibbons-Hawking Entropy}
\author{Gia Dvali } 
\affiliation{%
Arnold Sommerfeld Center, Ludwig-Maximilians-University,  Munich, Germany, 
}%
 \affiliation{%
Max-Planck-Institute for Physics, Munich, Germany
}%

\date{\today}

\begin{abstract} 
  We describe an attempt of string theoretic derivation of the Gibbons-Hawking entropy. 
  Despite not admitting a de Sitter vacuum, the string theory, 
  by the power of open-close correspondence,  captures  the Gibbons-Hawking entropy as the entropy of Chan-Paton species on a de Sitter-like state obtained via $D$-branes.   Moreover, this derivation sheds a new light at  the origin of the area-form, since  the  equality takes place for a critical  't Hooft coupling for which the species entropy of open strings 
 saturates the area-law unitarity bound.  

  \end{abstract}

\maketitle

\subsection{Introduction}

  Some time ago \cite{Gibbons:1977mu}
 Gibbons and Hawking pointed out that de Sitter space 
 possesses a microstate entropy that scales 
 as the area of the horizon, 
 \begin{equation} \label{SGH}
 S_{GH} \, \sim  \,  (R M_P)^2 \,,
 \end{equation} 
 where $R$ is the de Sitter curvature radius 
 and $M_P$ is the Planck mass.
  This expression  is very similar to the Bekenstein-Hawking 
  entropy of a black hole of equal radius \cite{Bekenstein:1973ur, Hawking:1974sw}. 
  
   While there is a full consensus about the correctness 
of  $S_{GH}$ entropy,  there exist no commonly accepted 
microscopic explanation for it. 
  It is natural to ask whether we can gain some insight 
  within string theoretic framework. 
  
  In this respect the situation  with de Sitter is rather peculiar
  and goes in contrast with black holes.   
   Namely, there exist a well-known string theoretic derivation 
  of Bekenstein entropy of an extremal black hole by
  Strominger and Vafa \cite{Strominger:1996sh}.  
  
   There also exist other stringy handles on black hole entropy.      
  For example, more recently, it became evident that the Bekenstein-Hawking entropy can be read out from the $S$-matrix elements of 
 multi-graviton and multi-closed string scattering processes at trans-Planckian center of mass  energies \cite{Dvali:2014ila, Addazi:2016ksu}.   Correspondingly,  such processes 
 can be interpreted as the $S$-matrix processes describing a black hole  formation at trans-Planckian energies 
\cite{tHooft:1987vrq, Amati:1987wq, Gross:1987kza}, 
in which the black hole is resolved as a multi-graviton 
coherent state  \cite{Dvali:2011aa}.  
  
  At the same time, we are not aware of any string theoretic derivation of Gibbons-Hawking entropy.  The goal of the present paper is to fill this omission.   
 We shall base our discussion on the arguments and the construction already 
 provided in \cite{Dvali:2021bsy}, where it was shown that via open-closed 
 duality the string theory carries a non-trivial 
 information about 
 the Gibbons-Hawking entropy of de Sitter.

   Let us fist notice that there are some fundamental differences between the ways in which black holes and de Sitter are treated by the string theory.     
       First, no de Sitter vacuum exist in string theory. Its absence 
   can be understood as a result of $S$-matrix on which the 
   formulation of string theory is intrinsically based \cite{Dvali:2020etd}.  Any eternally-inflating de Sitter-like universe makes a definition 
 of $S$-matrix impossible.  To our opinion, this feature is an intrinsic source of various issues with formulation of quantum gravity in de Sitter discussed by Witten \cite{Witten:2001kn}. 
 
  The absence of de Sitter $S$-matrix  justifies the earlier  
 proposals about incompatibility 
 of de Sitter with quantum gravity  \cite{Dvali:2013eja, Dvali:2014gua, Dvali:2017eba}. 
  It also fully resonates with the so-called 
  de Sitter conjecture \cite{Obied:2018sgi} and, 
  in the present context, even more closely  with the trans-Planckian censorship conjecture \cite{Bedroya:2019snp}.

 In contrast, as already remarked, black holes are in no conflict with $S$-matrix. 
  For example,  the unstable, non-extremal, black holes can be described as intermediate finite-lifetime resonances formed in an $S$-matrix process with trans-Planckian center of mass energies.    
  Alternatively, the stable black holes, such as the extremal 
  ones, can be viewed as legitimate asymptotic $S$-matrix states. 
  
   None of the above is possible for de Sitter.  Unlike 
   a black hole,  the de Sitter is an eternally-inflating state with no possibility of defining the $S$-matrix \cite{Dvali:2020etd}.      
    This difference can also be understood from the    
  fact that black holes can be formed 
  in a collision process of in-states  without any need of a
  vacuum energy. In contrast, de Sitter necessarily requires 
  an uniform classical source in form of a cosmological constant.

  However, the urgency for string-theoretic understanding 
 of Gibbons-Hawking entropy transcends the question 
 of the de Sitter  vacuum.  The reason is that the 
 Gibbons-Hawking entropy  is expected to be an intrinsic property of any slow-roll inflationary Hubble patch even with a moderate number  of e-foldings, not conflicting with the existence of an asymptotic $S$-matrix vacuum.

 In the present note, refining the earlier  
 work  \cite{Dvali:2021bsy},  we shall show that, despite not 
 supporting a de Sitter  vacuum, the string theory nevertheless 
 captures the origin of  Gibbons-Hawking entropy 
 thanks to the power of open-closed correspondence. 
    In the above paper it was observed that in the regime in which the system  admits a de Sitter like state, the Gibbons-Hawking entropy is accounted by the species entropy of 
 the open-string Chan-Paton modes. 
 We shall essentially repeat the construction of 
  \cite{Dvali:2021bsy}
   and attempt to give it a form of a full-fledged derivation.

\subsection{ A String Theoretic Construct} 

  Before moving to actual derivation,  we make some preparatory remarks and define some useful quantities. The two important 
parameters are: the species scale $M_{sp}$ and the species entropy $S_{sp}$.  
     
      The species scale sets an absolute cutoff on any 
  $d$-dimensional theory of gravity propagating $N_{sp} \gg 1$ particle species and is given by  \cite{Dvali:2007hz, Dvali:2007wp, Dvali:2008fd, Dvali:2008ec, Dvali:2009ks, Dvali:2010vm, Dvali:2012uq, Dvali:2020wqi},  \cite{Dvali:2021bsy},
  \begin{equation} \label{MSP}
    M_{sp} = \frac{M_{d}}{N_{sp}^{\frac{1}{d-2}}} \,,
  \end{equation} 
  where $M_d$ is the $d$-dimensional Planck mass. 
   The scale $1/M_{ps}$ is the lower bound on the 
   radius of a black hole beyond which the extrapolation
   of semi-classical features is impossible. 
  For example,  a would-be black hole of a smaller size 
 cannot maintain the thermal Hawking radiation, or else, 
 such a black hole would explode in less than the light-crossing 
 time, which would-be an absurd.  
 
 We must note that the species scale has perturbative manifestations  \cite{Dvali:2001gx, Veneziano:2001ah}, 
 however the power of the above 
 black hole arguments is that they are fully non-perturbative and therefore are insensitive to a possible fine-tuning or a resummation of the series.  
 More recent discussions of various non-perturbative manifestations of the species scale were given in   \cite{vandeHeisteeg:2022btw, vandeHeisteeg:2023ubh, vandeHeisteeg:2023dlw, Cribiori:2022nke}.

   Another important concept, on which we shall rely 
   heavily,  is  the ``species entropy", 
   $S_{sp}$,  which measures the capacity of the system to 
    store information in the microstates of 
    species \cite{Dvali:2008ec, Dvali:2010vm, Dvali:2020wqi},
    \cite{Dvali:2021bsy}. 
     For the species that are approximately degenerate in mass, 
   in a state with total occupation number $N$,       
     the species entropy  is given by the log of the binomial 
     coeficient \cite{Dvali:2020wqi}: 
     \begin{eqnarray} \label{EntSP}
S_{sp}  & = &   \ln\begin{pmatrix}
    N_{sp}  +  N   \\
      N_{sp}  
\end{pmatrix} \simeq  \\ \nonumber 
 & \simeq  & N_{sp}  \ln( (1+ N_{sp}/N)^{N/N_{sp}}
(1+N/N_{sp})) \,, 
\end{eqnarray}   
where in the last expression we used Stirling's approximation.
  As shown in  \cite{Dvali:2020wqi}, the 
 saturation of species entropy takes place for
 $N \sim N_{sp}$, which gives, 
  \begin{equation}  \label{Sspecies} 
 S_{sp} \sim N_{sp} \,.
\end{equation}

  The species entropy gives an alternative justification to the 
  statement that the cutoff is given by the species scale (\ref{MSP}).  
 Indeed, it has been shown \cite{Dvali:2008ec, Dvali:2010vm, Dvali:2020wqi} that for a would-be black hole 
  smaller than the species scale, $1/M_{sp}$, 
  the species entropy 
would exceed the Bekenstein-Hawking entropy, which is 
impossible. 
  
In our derivation, via open-closed correspondence, 
the Gibbons-Hawking entropy shall emerge as the species entropy of the open-strings  \cite{Dvali:2021bsy}.

 We start with an attempt of producing a de Sitter like state in string theory.  We shall use the standard  (anti)$D$-brane uplifting  mechanism put forward
in  \cite{Dvali:1998pa, Dvali:1999tq, Dvali:2001fw}. In fact,  we do not have much alternatives since, as already 
realized in the above work,  $D$-branes are the only known negative pressure 
sources in string theory.   

 The construction can be given in various 
 string dimensions. 
However, in order to avoid the secondary issues with technicalities of compactification,  we shall work in type {\it II}B in ten dimensions. 

 As usual, the two fundamental parameters are 
 the string coupling, $g_s$, and 
the string scale, $M_s$.  
 The ten-dimensional Planck mass, $M_{10}$, 
  is expressed  via  $M_s$ and $g_s$ as, 
   \begin{equation} \label{MsMp} 
   M_{10}^8  = \frac{M_s^8}{g_s^2} \,.
    \end{equation}      
    Throughout  the paper we shall omit the irrelevant numerical factors which can be easily restored.   
    
 Following \cite{Dvali:2021bsy}, we shall prepare 
a space-filling  background with 
 $n$  coincident $D_9$-branes and $n$ 
 anti-$\bar{D}_9$-branes.    
The world-volume (open string) physics of this system is 
rather well-understood  \cite{Srednicki:1998mq}. 
 The open string zero modes produce  $ N_{sp} \sim n^2$ 
light  Chan-Paton species which transform under the $U(n)\times U(n)$ gauge symmetry.

  As discussed in \cite{Dvali:2010vm}, such a background 
 has associated  ten-dimensional species scale 
  \begin{equation} \label{10sp} 
 M_{sp} \, = \, \frac{M_{10}}{n^{1/4}} \, = \frac{M_s}{(ng_s)^{\frac{1}{4}}}\,,
    \end{equation} 
 and the corresponding  
 species entropy  (\ref{EntSP}),  which at the 
 saturation point takes the value (\ref{Sspecies}),
 \begin{equation}  \label{SSP} 
 S_{sp} = n^2 \,.
\end{equation}  
   This entropy reflects the existence of 
  microstates with different excitation patterns  
   of the Chan-Paton species.  
   
  We shall be working in the 't Hooft  limit \cite{tHooft:1973alw} of  large $n$ and weak string coupling: 
   \begin{equation}  \label{Limit}
     n \rightarrow \infty\,,~ g_s  \rightarrow 0\,, 
   \end{equation}   
    while keeping the 't Hooft coupling,  $(ng_s)$, as well as 
    the string scale, $M_s$,  finite. We shall use the 
    't Hooft coupling as a control parameter, and shall   approach
  the critical value, $(ng_s) = 1$, from below. 
  
  Notice that this background generates a no-trivial scalar potential for the dilaton, thereby, likely making it time-dependent.
  We shall not enter in the discussion of 
  dilaton-stabilization \footnote{We thank Cumrun Vafa for 
  raising the question about dilaton stability.}. Instead, we shall 
  limit ourselves by restricting the time-scale of its potential 
  evolution.   
   The curvature scale of the dilaton effective potential  is set by the quantity, 
   \begin{equation} \label{dilaton} 
   M_s^2 (ng_s)\,.  
 \end{equation}  
   This makes the potential time-dependence of the dilaton sufficiently 
   slow for our arguments to go through. 
   We thus assume that on the time-scales of our interest, 
   dilaton is approximately frozen. 
      
 However,  the background contains an open-string tachyon, 
 with negative mass$^2$ of order $m^2_t \sim - M_s^2$.
 This reflects the instability  towards mutual annihilation of  $D$-branes and  anti-$D$-branes.
 This process can be described as the tachyon 
condensation \cite{Sen:1999mg}. 
 The classification is given in terms of $K$-theory
 \cite{Witten:1998cd}. 

 When tachyon condenses,  the  Chan-Paton species become massive.  A part of them gains the mass via the Higgs effect. 
Others become massive non-perturbatively.  
This is also obvious from the fact that the closed-string vacuum cannot support the open string excitations without $D$-branes.

 Now, the idea of  ``brane inflation" scenario
  \cite{Dvali:1998pa, Dvali:1999tq, Dvali:2001fw} is that, prior to tachyon condensation, the energy density coming from $D$-brane tensions acts as vacuum energy, creating a temporary de Sitter-like 
 inflationary state.  The tachyon condensation then 
 provides a graceful exit from inflation. 
  The duration of the inflationary state depends on the details of constructions in various dimensions to which we shall not
  enter.

 We shall rather focus on the state prior to tachyon condensation in which the tachyon is carefully tuned 
 in the position of unstable equilibrium, similarly to 
 a Higgs field on top of a ``Mexican hat" potential.

 For  a moment,  let us  mentally freeze the tachyon.   
  Then,  due to a positive vacuum energy density, 
  \begin{equation} \label{TD} 
  \Lambda_{10}  = n \frac{M_s^{10}}{g_s}  \,,
   \end{equation}
  coming from the collective tension of $D$-branes, classically, the system would produce a ten-dimensional de Sitter metric.  
  The corresponding de Sitter curvature radius would be
  \footnote{Comparing with (\ref{dilaton}), it is clear that 
  dilaton does not evolve faster that the Hubble time, which 
justifies the validity of our assumption.} 
 \begin{equation} \label{stringR} 
   R \,  = \, \sqrt{M_{10}^8/\Lambda_{10}} =  
   \frac{1}{\sqrt{ng_s} M_s} \,.
    \end{equation}  
  Therefore, the Gibbons-Hawking entropy
  would be equal to,  
  \begin{equation} \label{stringSGH} 
   S_{GH}  =  (R M_{10})^8 = \frac{1}{g_s^2(ng_s)^4} \,.
    \end{equation}  
    
   Thus, in a state with artificially frozen tachyon, the system 
   would posses two types of  entropies:  {\it  1)} The gravitational  
  Gibbons-Hawking entropy  (\ref{stringSGH}), 
  coming from closed strings;  and  {\it 2)} The Chan-Paton species entropy  
 (\ref{SSP}),  coming from open strings.   Their ratio scales as, 
   \begin{equation} \label{SPGH} 
   \frac{S_{sp}}{S_{GH}}  =  (ng_s)^6 \,.
    \end{equation}  
 This may look disappointing, since keeping in mind the open-closed correspondence, one could have hoped that the two entropies 
 would come out equal.  However, the situation is quite 
 the contrary. 
 
  In reality, none of the entropies are real, since the system is unstable on the time-scale less than the Hubble time. In order to talk about entropies, we must first reach the state in which the system is stabilized, at least for one Hubble time. 
  Only in such a state it makes sense to compare the 
  two entropies.

Now, following \cite{Dvali:2021bsy}, we can try to self-consistently stabilize the tachyon
using a general mechanism  discussed earlier in  \cite{Dvali:1999tq}.
  We proceed in the following way. 
  Instead of Chan-Paton vacuum, we consider a state in which the  species are excited. Such excitations generate 
  a positive contribution into the mass$^2$ of the tachyon field
  and can potentially stabilize it.  
  
  In fact, in a would-be de Sitter, the Chan-Paton species 
 would be automatically excited due to the Gibbons-Hawking 
 temperature of de Sitter,
 \begin{equation} \label{TGH} 
  T_{GH}  =  \frac{1}{R} \,.
    \end{equation}     
 Thus, we can proceed in the following manner. We can assume that tachyon has been stabilized and check the self-consistency 
 of this assumption by verifying the thermal
 contribution to its mass from the Chan-Paton species.

In a would-be de Sitter, the effective mass-square of the tachyon 
would receive the following 
 positive contribution from the coupling to the thermal bath of 
 Chan-Paton species  at temperature $T_{GH}$
 \footnote{Notice that a one-loop correction from stringy version 
 of Coleman-Weinberg potential shall not overpower the 
 thermal correction, if we use as a rough estimate 
 the calculation of \cite{Dvali:1994ms} with multiplicity of 
 species $n^2$ and the coupling $g_s^2$.}  
\begin{equation} \label{DeltaM} 
  \Delta m^2_{t} = 
  \frac{(ng_s)^2}{R^2} =  (ng_s)^3 M_s^2 \,.
    \end{equation}     
 This contribution  would balance 
 the  zero-temperature negative 
 mass-square of the tachyon, provided the parameters satisfy,  
 \begin{equation} \label{CPLimit} 
  (ng_s) \sim 1 \,.
    \end{equation}  
  In this case the stack of branes could be stabilized.\\
  
  This mechanism of potential stabilization of $D$-branes 
  on top of each other was previously considered 
  in  \cite{Dvali:1999tq} and can be viewed as the effect of  restoration of $U(n)\times U(n)$ symmetry at
  non-zero temperature.  
  
   The remarkable thing is that at the critical point 
 (\ref{CPLimit}) the Gibbons-Hawking entropy 
 becomes equal to the entropy of species (\ref{SPGH}), as 
 this would be expected from the open-closed correspondence.
 
  The presented reasoning can be viewed 
  as the microscopic understanding of Gibbons-Hawking 
  entropy in terms of the entropy of Chan-Paton species.  
     
 One may complain that there is a potential loophole  
  because one could seemingly  violate the equality 
  between the two entropies 
 by  taking
 \begin{equation} \label{ng1}
   (n g_s) \gg 1\,.
  \end{equation}  
   This is however not possible, since it would 
  put our calculation out of the domain of validity. 
  This is signalled by several factors.

  First, it was already noticed in  \cite{Dvali:2010vm} that for (\ref{ng1}), the ten-dimensional species scale 
  (\ref{10sp}) 
  which sets the absolute cutoff of effective low energy theory, 
 would drop below the string scale $M_s$. 
 This is not possible,  since the cutoff in string theory is 
 the string scale. 
    
    Simultaneously, the de Sitter curvature radius 
    $R$  (\ref{stringR}) would become shorter than the string length $1/M_s$. 
    This would signal that  any geometric notion of 
 de Sitter becomes non-existent. 
 
  In particular, the Gibbons-Hawking temperature  (\ref{TGH}) 
   of a would-be de Sitter state exceeds the string scale and 
  the Hagedorn effects \cite{Hagedorn:1965st}
must be taken into account \cite{Bowick:1992qu}. 
  At this point various stringy effects, such as Atick-Witten phase transition \cite{Atick:1988si},
 also enter \footnote{Various discussion can be found in    
 \cite{Alvarez:1986sj}, \cite{Dienes:2012dc}, 
 and references therein.}. 
 

 The important message that we deduce from  our analysis 
 is that in the unique regime (\ref{CPLimit}), in which 
 the system can potentially deliver a de Sitter-like state, 
 the Gibbons-Hawking entropy (\ref{stringSGH})   is understood
 as the species entropy (\ref{EntSP}) of the Chan-Paton species.
  At the critical point (\ref{CPLimit}),  the two entropies are equal and they satisfy: 
     \begin{equation} \label{CPGH}
   S_{GH} = S_{sp} = n^2 = \frac{1}{g_s^2} \,.
  \end{equation}     
   
    Thus, by the power of open-closed correspondence, 
    string theory offers a microscopic understanding 
    of Gibbons-Hawking entropy in terms of 
  species entropy of open strings.    
      
 \subsection{Species Origin of the Area-Law} 
 
  With the equation (\ref{CPGH}) our task of 
  string theoretic derivation of Gibbons-Hawking entropy 
  is complete.   What comes next is the interpretational part. 
    Although this interpretation changes nothing in the above derivation, it is nevertheless physically important, as it opens 
    up a new view on open-closed correspondence. 
    
  Namely, we would like to give some physical intuition 
  to a ``miracle" of obtaining an intrinsically-geometric 
  area-form of Gibbons-Hawking entropy (\ref{stringSGH}), from a seemingly-non-geometric concept of open string species entropy (\ref{EntSP}).   

   In  fact, the story goes beyond the usual view of ``holography" in which a geometric area-count of  
the entropy is an exclusive feature of the gravitational part of the theory (for a review, see \cite{Bousso:2002ju}). 
 
 Instead, as we shall see, both entropies, $S_{sp}$ and $S_{GH}$,  independently have very clear geometric area-forms.  This feature is fundamentally linked 
  with the criticality of the 't Hooft  coupling (\ref{CPLimit}).
  Moreover, at the same time, for the critical coupling 
  the two entropies are equal (\ref{CPGH}).    That is, 
 in certain sense,  the open-closed correspondence is responsible 
not for the area-forms of the two entropies {\it per se}, but for their equalities at the critical 't Hooft  coupling (\ref{CPLimit}).   

   The key point is that, irrespective of gravity, the species microstate 
   entropy does have a geometric meaning in form of 
   the following universal bound    
  on the 
 microstate degeneracy \cite{Dvali:2020wqi}:  
  \begin{equation}   \label{GoldE}
   S_{max} = (Rf)^{d-2} \,.
 \end{equation} 
 This bound  implies that, within  the validity of a given  description, the maximal microstate entropy attained by a $d$-dimensional system of a radius $R$ is equal to its  surface area in units of the scale of a  spontaneously broken Poincare symmetry, $f$. 
 
The validity domain is defined in the usual quantum field theoretic 
(or string theoretic) sense  as the requirement 
of a weak coupling of the relevant degrees of freedom. 
In particular, the saturation of the bound (\ref{GoldE}) is correlated with 
  the saturation of unitarity by a set of scattering amplitudes \cite{Dvali:2020wqi}.   
  
  The bound (\ref{GoldE}) is universal and is independent of gravity,  as long as the system can be defined as 
  a state on an asymptotic vacuum of Minkowski. 
   In the present case, this condition is satisfied,   since our de Sitter-like state is not eternal and is going to asymptote 
   into a ten-dimensional Minkowski vacuum.  
    We can therefore classify the states according to the Poincare symmetry  of that vacuum.

  Here, we shall not enter in the justification of the bound 
  (\ref{GoldE}) for generic systems, which can be found in 
  \cite{Dvali:2020wqi} and in many related papers 
  (for an incomplete list, see, \cite{Dvali:2019jjw, Dvali:2019ulr, Dvali:2021ooc, Dvali:2021jto, Dvali:2021rlf, Dvali:2021tez, Dvali:2021ofp}). 
    Instead, we shall limit ourselves by 
   providing an intuition  for the formula 
  (\ref{GoldE}) specifically for the species entropy, again following  \cite{Dvali:2020wqi}. 
  
   Let us imagine a $d$-dimensional system of radius $R$ with a total occupation number $N$ of massless particle species.  
   Such a system breaks the Poincare symmetry at  
   the scale 
   \begin{equation} \label{fd}
  f = \frac{N^{\frac{1}{d-2}}}{R}\,. 
  \end{equation} 
   This is easy to understand from the fact that
   the quantity $1/N$ is the dimensionless coupling of  
  the corresponding Goldstone mode.    
   
   At the same time, the species entropy
   (\ref{EntSP}) at the saturation point 
 is given by  $N = N_{sp}$.  This immediately shows 
 that (\ref{Sspecies}) is given by (\ref{GoldE}). 
  Thus, even in the flat space, the species entropy, at its saturation point,  carries a very clear area-law  meaning.

    Applied to our $D$-brane construction,  
  it is easy to see that at the critical point 
  (\ref{CPLimit}) the scale of spontaneous breaking of the 
  ten-dimensional Poincare invariance 
   is the ten-dimensional Planck scale,
   \begin{equation} \label{fM10}
   f = \frac{n^{\frac{1}{4}}}{R} = M_{10}\,.  
  \end{equation}  
  Correspondingly, the entropy 
  of Chan-Paton species (\ref{SSP}), which was  
already shown   to be equal to Gibbons-Hawking entropy
(\ref{CPGH}), at the same time,  saturates the
   area-law bound (\ref{GoldE}) with $f = M_{10}$. 
   Obviously, for $f= M_{10}$, the equation  (\ref{GoldE}) matches the  general Gibbons-Hawking expression (\ref{stringSGH}).

   We thus see that the open-closed correspondence allows to understand the Gibbons-Hawking entropy,
  and most importantly its area-form,  
    as  the saturation of the universal bound (\ref{GoldE}) by the Chan-Paton species entropy (\ref{SSP}) at the critical 't Hooft coupling  (\ref{CPLimit}).

 \subsection{Some Implications} 
 
  Here we briefly discuss some  implications of our results. 
 
 \subsubsection{Glimpses of de Sitter Memory Burden} 
  
 The above construction also allows us to catch glimpses of
 a de Sitter  ``memory-burden"  effect \cite{Dvali:2018ytn}.  
  The essence of the memory burden phenomenon
   \cite{Dvali:2018xpy, Dvali:2018ytn, Dvali:2020wft, Dvali:2024hsb}  is that the 
 load of information tends to stabilize the system. 
  The effect is especially prominent in systems 
  of maximal capacity of information (memory) storage, in particular,  
  the systems that saturate the bound (\ref{GoldE})
 on  the microstate degeneracy, so-called ``saturons". 
  
  In such a system the information is stored in a set of degrees of freedom called ``memory modes".  Lower is the excitation energy cost of the memory mode, more efficient is the information storage 
capacity of the system. 
  
   The special feature of enhanced memory systems is that 
   they possess a critical state in which many 
   species of nearly-gapless memory modes emerge.  This criticality can be parameterized by a control order parameter called a ``master mode". 
  
   The load of information corresponds to an excitation pattern 
   of the memory modes. This stabilizes the system 
   in the state in which these modes are least costly in energy. 
 
   Since the effect is universally present in systems of high microstate degeneracy,  it is expected to be 
  operative in black holes  \cite{Dvali:2018xpy, Dvali:2020wft, 
 Alexandre:2024nuo, Thoss:2024hsr}  and in de Sitter  \cite{Dvali:2018ytn}. 
 
 In case of a black hole, the memory burden is expected 
 to slow down the decay after emission of about  a half of the mass.      
 The  localized objects such as 
 black holes   \cite{Dvali:2018xpy, Dvali:2020wft, Dvali:2024hsb}  
 or solitons \cite{Dvali:2021tez, Dvali:2024hsb}, when  burdened 
by memory,  are relatively straightforward to be visualized. 
  
  However, the analogous question for de Sitter is more   
  profound \cite{Dvali:2018ytn}. In particular, this is the case due to absence of an external observer. 
        
   In the present construction the meaning of the  memory burden effect  in a de Sitter-like state  becomes rather explicit.
  Indeed, the memory modes are the 
 Chan-Paton species, whereas tachyon is a master mode.  When tachyon is on top of the Mexican hat, the memory modes are gapless. 
 If tachyon condenses, system moves to the closed-string vacuum and the memory modes gain large mass 
 gaps.  For  excited memory modes,
 this would be very costly in energy.      
 Correspondingly, the excitations of Chan-Paton modes 
tend to create an energy barrier and stabilize the tachyon on top of the hill.   
  
    In this way, the string theory allows us to catch  
  a glimpse  of a de Sitter  burdened by memory: it is a state at critical  't Hooft coupling  (\ref{CPLimit})  with quantum stringy effects becoming as important as the classical ones.  \\

 \subsubsection{Quantum Breaking of de Sitter} 
 
 From  the $S$-matrix perspective \cite{Dvali:2020etd},  the de Sitter 
 is expected to be describable as a coherent state of gravitons 
 constructed on top of an asymptotic $S$-matrix vacuum of Minkowski \cite{Dvali:2013eja, Dvali:2014gua, Dvali:2017eba} (for a BRST-invariant construction, see 
 \cite{Berezhiani:2021zst}) 
\footnote{ Interestingly, this  picture passes an immediate consistency test of reproducing the Gibbons-Hawking 
entropy (\ref{stringSGH}) as a particular case 
of saturation of the bound (\ref{GoldE}). 
 Indeed, since the asymptotic vacuum is Minkowski, the notion of 
 spontaneous breaking of Poincare symmetry by the  de Sitter 
 coherent state is well-defined and the scale of breaking is the Planck mass $f = M_d$, which again reproduces 
(\ref{stringSGH}) from (\ref{GoldE}).}.
  This picture comes with an  accompanying phenomenon of ``quantum-breaking".  Its essence is that the quantum evolution of the coherent state of gravitons, after 
a so-called quantum break-time,  $t_Q$,   
fully departs from the classical one. 
  
   In  \cite{Dvali:2013eja} the quantum break-time from the inner entanglement of the graviton state was suggested to be $t_Q \sim R \ln(S_{GH})$.  This is very similar to the time-scale given by  the trans-Planckian  censorship  conjecture \cite{Bedroya:2019snp}.  
 For the present stringy construction  $t_Q$ 
translates as \cite{Dvali:2020etd} $t_Q \sim R \ln(1/g_s^2)$.    
 Needless to say, a more precise stringy computation of the log-factor would be highly informative. 
 
 \subsection{Outlook}
 
  In this note we offered a string theoretic view at the microscopic 
origin of  the Gibbons-Hawking entropy. 
  In the limit of critical 't Hooft coupling in which a  
set of $D-\bar{D}$-branes delivers a de Sitter-like state, 
the entropy of open string 
Chan-Paton species reproduces the gravitational 
Gibbons-Hawking entropy.  
  The crux of it all is in universal features of species entropy
  as well as in  open-close correspondence of string
  theory.  This allows for understanding of intrinsically gravitational  entropy of closed strings  in terms of  non-gravitational entropy of the Chan-Patton species.

    Remarkably,  the underlying physics of this connection goes beyond the ordinary view on ``holography" in which 
 the geometric area-form is the exceptional property of only 
 one side of the description.  
    Instead,  the non-gravitational 
species  entropy, $S_{sp}$,  at the 
saturation of unitarity, independently, has the area-form (\ref{GoldE}) controlled by the strength of spontaneous 
breaking of the Poincare symmetry. 
   At the critical value (\ref{CPLimit}) of the stringy 't Hooft coupling,  the two entropies become equal (\ref{CPGH}).

        The following comments are 
in order.  
  First, of course, we expect corrections from higher order operators \cite{CV}.  It is important to understand how they correct  the equation (\ref{CPGH}).  
  
    Secondly,  the fact that our construction is valid up to a critical coupling (\ref{CPLimit}),  
  prompts the question  whether an analog of 
 Horowitz-Polchinski string-black hole correspondence  \cite{Horowitz:1996nw}  can be defined for the 
 de Sitter case \cite{CV}.   
In certain sense, the presented construction indicates 
a path in this direction via providing a correspondence
 between the entropies of open-strings and Gibbons-Hawking.

  Finally, it would be proper  to understand a potential connection  with AdS/CFT correspondence \cite{Maldacena:1997re, Gubser:1998bc, Witten:1998qj}.
 \\

 {\bf Acknowledgements} \\
 
    We thank organizers of Lemaitre Conference 2024
where this work was presented. 
 It is pleasure to thank Cumrun Vafa for discussions.  

 This work was supported in part by the Humboldt Foundation under the Humboldt Professorship Award, by the European Research Council Gravities Horizon Grant AO number: 850 173-6, by the Deutsche Forschungsgemeinschaft (DFG, German Research Foundation) under Germany’s Excellence Strategy - EXC-2111 - 390814868, Germany’s Excellence Strategy under Excellence Cluster Origins 
EXC 2094 – 390783311. The work of AK and OS was partially supported by the Australian Research Council under the Discovery Projects grants DP210101636 and DP220101721.     \\

\noindent {\bf Disclaimer:} Funded by the European Union. Views and opinions expressed are however those of the authors only and do not necessarily reflect those of the European Union or European Research Council. Neither the European Union nor the granting authority can be held responsible for them.\\

\end{document}